\input{aipcheck}

\documentclass[
    ,final            
  ]
  {aipproc}

\layoutstyle{8x11double}

\newcommand {\bc}{\begin{center}}
\newcommand {\ec}{\end{center}}
\newcommand {\bea}{\begin{eqnarray}}
\newcommand {\eea}{\end{eqnarray}}
\newcommand {\be}{\begin{equation}}
\newcommand {\ee}{\end{equation}}

\def\lsim{\mathrel{\rlap{\lower4pt\hbox{\hskip1pt$\sim$}}
    \raise1pt\hbox{$<$}}}               
\def\gsim{\mathrel{\rlap{\lower4pt\hbox{\hskip1pt$\sim$}}
    \raise1pt\hbox{$>$}}}                
\usepackage{graphicx}
\begin{document}

\title{Elliptic flow and nearly perfect fluidity in dilute Fermi gases}


\author{Thomas Sch\"afer}{
  address={Department of Physics, North Carolina State University,
           Raleigh, NC 27695}
}

\begin{abstract}
In this contribution we summarize recent progress in understanding 
the shear viscosity of strongly correlated dilute Fermi gases. We
discuss predictions from kinetic theory, and show how these
predictions can be tested using recent experimental data on elliptic 
flow. We find agreement between theory and experiments in the high 
temperature regime $T\gg T_F$, where $T_F$ is the the temperature 
where quantum degeneracy effects become important. In the low temperature 
regime, $T\sim T_F$, the strongest constraints on the shear viscosity
come from experimental studies of the damping of collective modes. 
These experiments indicate that $\eta/s\lsim 0.5\hbar/k_B$, where 
$\eta$ is the shear viscosity and $s$ is the entropy density. 
\end{abstract}

\maketitle


\section{Introduction}

 Experiments carried out at the relativistic heavy ion collider
(RHIC) have demonstrated that the quark gluon plasma is a very
good fluid \cite{Adler:2003kt,Back:2004mh,Adams:2004bi}. A 
quantitative analysis of the observed elliptic flow in the 
framework of viscous relativistic hydrodynamics shows that 
\cite{Dusling:2007gi,Romatschke:2007mq}
\be 
\frac{\eta}{s}\lsim 0.5 \, \frac{\hbar}{k_B}\, 
\ee
where $\eta$ is the shear viscosity and $s$ is the entropy density. 
The best fit to the data corresponds to even smaller values, 
$\eta/s\simeq (0.1-0.2)\hbar/k_B$. This number is close to a 
proposed lower bound, $\eta/s=\hbar/(4\pi k_B)$, on the ratio 
of shear viscosity to entropy density that is saturated in the 
strong coupling limit of a large class of field theories that 
can be analyzed using the AdS/CFT correspondence 
\cite{Policastro:2001yc,Kovtun:2004de}. 

 The RHIC results raise the question whether nearly perfect 
fluidity is a phenomenon that is specific to relativistic 
gauge theories, or whether it is a more general effect that 
also appears in other strongly correlated quantum fluids. If 
that is the case we may also ask whether there are universal 
aspects of nearly perfect fluidity that go beyond the relation 
$\eta/s \simeq \hbar/(4\pi k_B)$. Other universal features
could include bounds on other transport coefficients like 
diffusion constants or relaxation times, or constraints 
on spectral properties of the theory. In the context of 
interpreting the RHIC results we are particularly interested
in the question whether nearly perfect fluidity is consistent
with a quasi-particle description of the fluid. 

 Cold, dilute Fermi gases in which the interaction between the 
atoms can be tuned using an external magnetic field provide a 
new paradigm for strongly correlated quantum fluids 
\cite{Bloch:2007,Giorgini:2008,Schafer:2009dj}. These systems 
have been realized experimentally using optically trapped alkali 
atoms such as $^6$Li and $^{40}$K. These two atoms are fermions because
they posses is a single valence electron and the nuclear spin is integer.
When a dilute gas of alkali atoms is cooled to very low temperature, 
we can view the atoms as pointlike particles interacting via interatomic 
potentials which depend on the hyperfine quantum numbers of the 
valence electron. A Feshbach resonance arises if a molecular 
bound state in a ``closed'' hyperfine channel crosses the 
threshold of an energetically lower ``open'' channel. Because the 
magnetic moments of the states in the open and closed channel are 
in general different, Feshbach resonances can be tuned using an 
applied magnetic field. At resonance the two-body scattering length 
in the open channel diverges, and the cross section $\sigma$ is 
limited only by unitarity, $\sigma(k) = 4\pi/k^2$ where $k$ is 
the relative momentum. 

 In the unitarity limit details of the microscopic interaction are 
irrelevant, and the system displays universal properties. The 
dilute Fermi gas can be described by the effective lagrangian
\be 
\label{l_4f}
{\cal L} = \psi^\dagger \left( i\partial_0 +
 \frac{{\bf \nabla}^2}{2m} \right) \psi 
 - \frac{C_0}{2} \left(\psi^\dagger \psi\right)^2 ,
\ee
were $\psi$ is a two-component fermion field with mass $m$. The 
coupling constant $C_0$ is related to the $s$-wave scattering length 
$a$. The precise form of this relation depends on the regularization 
scheme. In dimensional regularization one finds $C_0 = (4\pi a)/m$. 
Terms with more derivatives or higher powers of the field are related 
to effective range corrections, higher partial waves, and many-body 
forces. All of these terms are irrelevant in the universal limit. The 
two-body scattering cross section is 
\be 
\sigma(k) = \frac{4\pi a^2}{1+a^2k^2} \, . 
\ee
In the limit $a\to\infty$ the theory is parameter free, strongly 
interacting and scale as well as conformally invariant.

\section{Thermodynamics}
\label{sec_thermo}

 In the weak coupling limit $a\to 0$ the equation of state of the 
dilute Fermi gas is that of a free two-component Fermi gas, 
\be
\label{eos_0}
 P_0(\mu,T) = -2T\lambda_{\it dB}^{-3} {\it Li}_{5/2}(-\zeta^{-1})\, ,
\ee
where $\lambda_{\it dB}=[(2\pi)/(mT)]^{1/2}$ is the de Broglie
wave length, ${\it Li}_{\alpha}(x)$ is the Polylogarithm function, 
and $\zeta=\exp(-\mu/T)$ is the fugacity. At unitarity scale 
invariance implies that the equation of state is of the form
\be 
\label{eos}
 P(\mu,T) = \frac{h(\zeta)}{2}P_0(\mu,T)\, ,
\ee
where $h(\zeta)$ is a universal function. This function can be 
calculated using the Virial expansion at high temperature, $\zeta\gg 1$, 
but it is a non-perturbative quantity at low temperature. The 
equation of state is known to about 10\% from experimental 
measurements \cite{Luo:2008,Chevy:2009,Horikoshi:2010} and quantum 
Monte Carlo simulations \cite{Lee:2005it,Burovski:2006,Bulgac:2008zz}.
At zero temperature $h(0)/2=\xi^{-3/2}$, where $\xi\simeq 0.4$ is 
known as the Bertsch parameter. Scale invariance also implies that 
\be 
\label{eos_2}
 P  = \frac{2}{3} \varepsilon \, , 
\ee
where $\varepsilon$ is the energy density. At low temperature the 
attractive interaction between the fermions leads to superfluidity.
The critical temperature is $T_c\simeq 0.15 E_F$ \cite{Burovski:2006}, 
where $E_F$ is the Fermi energy. The Fermi energy of the interacting 
gas is defined by $E_F=k_F^2/(2m)$ where $k_F=(3\pi^2 n)^{1/3}$ is 
the Fermi momentum and $n$ the fermion density. 

\section{Hydrodynamics}
\label{sec_hydro}

 At large distances and long times deviations from equilibrium
are described by hydrodynamics. For simplicity we will consider 
the unitary Fermi gas in the normal phase. In that case there 
are five hydrodynamic variables, the mass density $\rho=mn$, the flow 
velocity $\vec{v}$, and the energy density ${\cal E}$. These 
variables satisfy five hydrodynamic equations, the continuity 
equation, the Navier-Stokes equation, and the equation of 
energy conservation,   
\bea
\label{hydro1}
\frac{\partial \rho}{\partial t} 
   + \vec{\nabla}\cdot\left(\rho\vec{v}\right)  &=& 0 , \\
\label{hydro2}
 \frac{\partial (\rho v_i)}{\partial t}  
   + \nabla_j\Pi_{ij} &=& 0, \\
\label{hydro3}
 \frac{\partial {\cal E}}{\partial t} 
 \; +\; \nabla_i j_i^{\;\epsilon} &=& 0 .  
\eea 
The total energy density is the sum of the internal energy
density and kinetic energy density, ${\cal E}=\varepsilon+\frac{1}{2}
\rho v^2$. These equations close once we supply constitutive relations 
for the stress tensor $\Pi_{ij}$ and the energy current $j_i^{\;\epsilon}$ 
as well as an equation of state. As explained in the previous section
the equation of state is $P=\frac{2}{3}\varepsilon$. The stress tensor 
is given by 
\be 
 \Pi_{ij} = \rho v_i v_j + P\delta_{ij}+ \delta \Pi_{ij}\, ,
\ee
where $\delta\Pi_{ij}$ is the dissipative part. The dissipative
contribution to the stress tensor is $\delta\Pi_{ij}=-\eta
\sigma_{ij}$ with
\be 
\label{sig_NS}
 \sigma_{ij} = \left(\nabla_i v_j +\nabla_j v_i 
  -\frac{2}{3}\delta_{ij}(\nabla_k v_k)\right)\, ,
\ee
where $\eta$ is the shear viscosity and we have used the fact that 
the bulk viscosity of the unitary Fermi gas is zero \cite{Son:2005tj}. 
The energy current is 
\be
j_i^{\;\epsilon} = v_iw+\delta j_i^{\;\epsilon}\, ,
\ee 
where $w=\varepsilon+P$ is the enthalpy density. The dissipative 
energy current is 
\be 
\delta j_i^{\;\epsilon} = \delta\Pi_{ij} v_j - 
\kappa \nabla_i T\, , 
\ee
where $T$ is the temperature and $\kappa$ is the thermal conductivity. 
We note that the temperature $T=T(n,P)$ is a function of the density 
$n=\rho/m$ and the pressure. In order to determine $T$ we need the 
equation of state in the form $P=P(n,T)$. Universality implies that 
$P(n,T)=m^{-1}n^{5/3}f(mT/n^{2/3})$ where $f(x)$ is a universal function 
that is related to the function $h(\zeta)$ defined in equ.~(\ref{eos}).
The situation simplifies in the high temperature limit where $P=nT$. 
Universality also restrict the dependence of the shear viscosity and 
thermal conductivity on the density and the temperature. We can write 
\bea 
\label{alpha_n}
\eta(n,T) &=& \alpha_n\left(\frac{mT}{n^{2/3}}\right)\, n \, , \\
\kappa(n,T) &=& \sigma_n\left(\frac{mT}{n^{2/3}}\right)\, \frac{n}{m} \, , 
\eea
where $\alpha_n(y)$ and $\sigma_n(y)$ are universal functions of
$y=mT/n^{2/3}$. The relative importance of thermal and momentum
diffusion can be characterized in terms of a dimensionless ratio
known an the Prandtl number, ${\it Pr}=c_p\eta/(\rho\kappa)$, where
$c_p$ is the specific heat at constant pressure. In the high 
temperature limit $c_p=\rho/m$ and ${\it Pr}=\alpha_n/\sigma_n$. 
Kinetic theory predicts that in this limit ${\it Pr}=2/3$
\cite{Braby:2010ec}.

\section{Kinetic theory}
\label{sec_kin}

 Near $T_c$ the transport coefficients $\eta$ and $\kappa$ are 
non-perturbative quantities that have to be extracted from 
experiment or computed in quantum Monte Carlo calculations. 
At high temperature (and at very low temperature, $T\ll T_c$, 
see \cite{Rupak:2007vp}) transport coefficients can be computed
in kinetic theory. The shear viscosity was first computed in 
\cite{Bruun:2005}. Here we will follow the recent work 
\cite{Braby:2010xx} which also considers the frequency 
dependence of the shear viscosity.

 In kinetic theory the stress tensor is given by 
\be
\label{T_ij_kin}
\Pi_{ij} = 2\int \frac{d^3p}{(2\pi)^3} 
     \frac{p^i p^j}{m}f_p\, ,
\ee
where $f_p=f(t,x,p)$ is the distribution function of fermion 
quasi-particles and the factor 2 is the spin degeneracy. The
equation of motion for $f_p$ is the Boltzmann equation. In 
order to extract the shear viscosity it is useful to consider
the Boltzmann equation in a background gravitational field. 
In this setting correlation functions of the stress tensor
can be determined by computing variational derivatives with 
respect to the background metric. The non-relativistic limit
of the Boltzmann equation in a curved space characterized
by the metric $g_{ij}$ is 
\bea
\label{BE_nr}
\lefteqn{ \Big( \frac{\partial}{\partial t}
                  + \frac{p^i}{m}\frac{\partial}{\partial x^i} 
  - \Big(  g^{il}\dot{g}_{lj}p^j} &&    \nonumber \\
 &&  \mbox{}+ \Gamma^{i}_{jk}\frac{p^{j}p^{k}}{m}\Big) 
       \frac{\partial}{\partial p^{i}}\Big) 
    f(t,x,\mathbf{p}) = C[f]\, ,
\eea
where $\Gamma^{i}_{jk}$ is the Christoffel symbol and $C[f]$ is the 
collision integral. We consider small deviations from equilibrium 
and write $f=f_{0}+\delta f$ with $f_{0}(\mathbf{p})=f_{0}(p^{i}p^{j}
g_{ij}/(2mT))$. We also write $g_{ij}=\delta_{ij}+h_{ij}$ and linearize 
in $h_{ij}$ and $\delta f$. We get
\be
\label{boltzmann_linear}
\left(\frac{\partial}{\partial t}
  +\frac{p^i}{m}\frac{\partial}{\partial x^i}\right)\delta f
   +\frac{f_{0}(1-f_{0})}{2mT} p^{i}p^{j}\dot{h}_{ij}
   = C[\delta f]\, . 
\ee
This equation can be solved by making an ansatz for $\delta f$. 
We go to Fourier space and and write
\be
\label{del_f_ans}
\delta f(\omega,k,p)=\omega f_{0}(1-f_{0}) \frac{p^i p^j}{2mT}
     \frac{\xi_T h^T_{ij}+\xi_L h^L_{ij} }
          {\omega-v_{p}\cdot k+ i\epsilon}
\ee
where $\xi^{T,L}=\xi^{T,L}(\omega,k)$, $h_{ij}^{T,L}$ are 
the traceless/trace parts of $h_{ij}$, and $v^i_p=p^i/m$ is the 
quasi-particle velocity. Inserting this ansatz into the 
Boltzmann equation we can solve for $\xi^{T,L}$ and then 
compute $\delta f$ and $T_{ij}$. Matching the result to 
hydrodynamics determines the shear viscosity. In the limit 
$k\to 0$ this can be done analytically. The zero frequency
limit of the shear viscosity is 
\be
\label{eta_0}
\eta = \frac{15 (mT)^{3/2}}{32 \sqrt{\pi}} 
   \left\{
   \begin{array}{cl}
   1 & a\to \infty \\
   1/(3mTa^2) & a \to 0
   \end{array}\right. \,  . 
\ee
We observe that the shear viscosity is large in the weak coupling 
limit $a\to 0$, and that $\eta$ decreases as the scattering length 
is increased. We note, however, that the shear viscosity saturates
when the scattering length becomes comparable to the de Broglie wave 
length $a\sim \lambda_{\it dB}\sim (mT)^{-1/2}$. In this limit $\eta$ 
only depends on $\lambda_{\it dB}$ but not on the density or the 
scattering length. The frequency dependence of the shear viscosity is
\be
\label{eta_w}
\eta(\omega) = \frac{\eta}{1+\omega^2\tau^2}
\ee
where $\tau = (3\eta)/(2\varepsilon)$ is the relaxation time
\cite{Bruun:2007}. The relaxation time controls the time scale
over which the viscous stress tensor relaxes to the Navier-Stokes
form given in equ.~(\ref{sig_NS}). We observe that relaxation is fast 
in the limit where the viscosity is small. We also observe 
that the viscosity satisfies a sum rule which only depends
on thermodynamic quantities, 
\be
\label{sum_rule}
\frac{1}{\pi}\int_0^{\infty} d \omega\, \eta(\omega) 
  = \frac{\varepsilon}{3}\, . 
\ee
A modified version of this sum rule which contains an extra
short time (high frequency) contribution can be proven in the 
full quantum theory \cite{Taylor:2010ju}.

\begin{figure}[t]
\includegraphics[width=.48\textwidth]{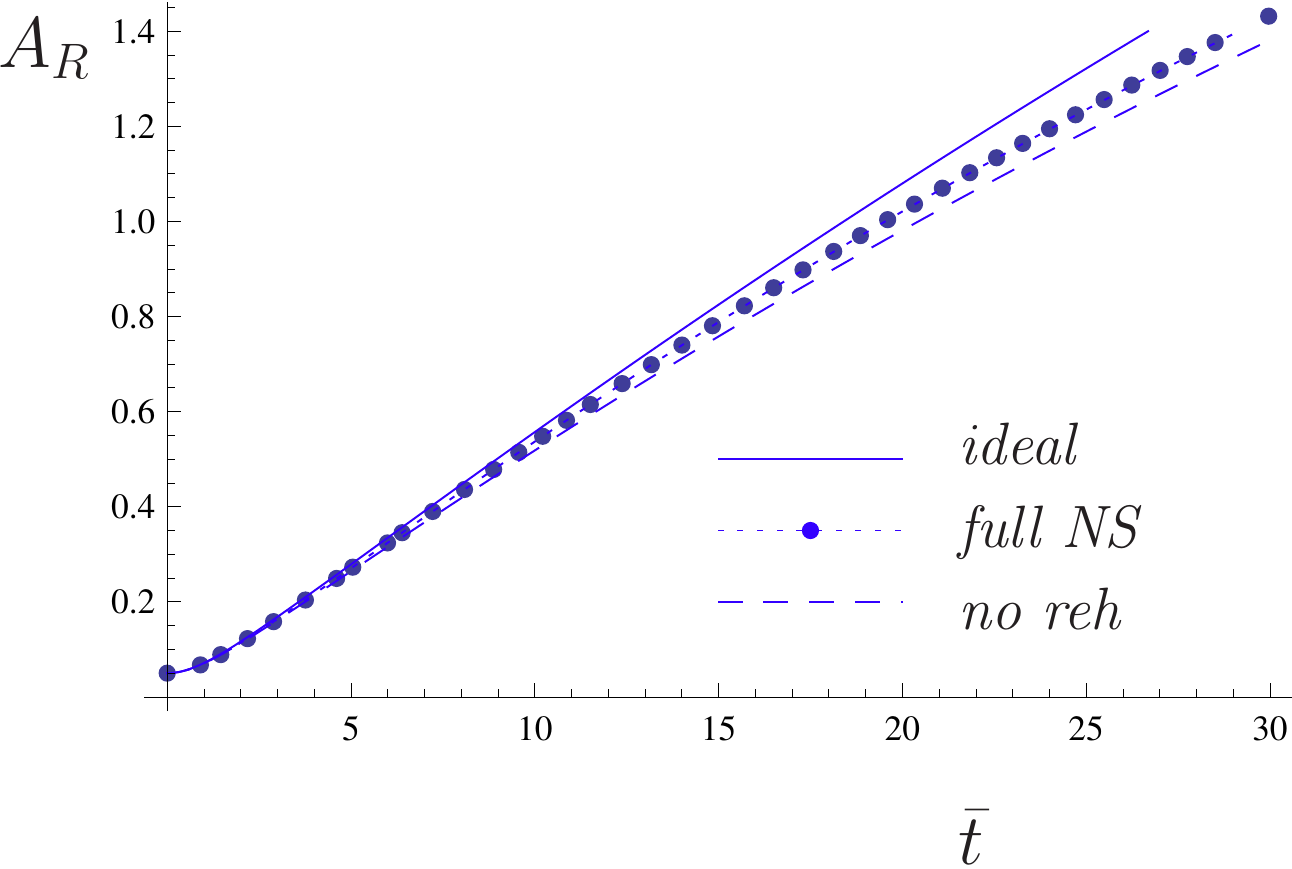}
\caption{\label{fig_A_R}
Time evolution of the aspect ratio $A_R$ of a deformed cloud as a 
function of the dimensionless time variable $\bar{t}=\omega_\perp t$, 
where $\omega_\perp$ is the harmonic oscillator constant of the 
transverse confining potential. The solid lines show the analytical 
result for the ideal evolution, the dashed lines correspond to 
the dissipative solution without reheating, and the dotted line 
shows the solution of equ.~(\ref{ns_for_1}-\ref{ns_for_4}). The 
points are from a numerical solution of the hydrodynamic equations.
The viscosity parameter is $\beta=0.066$, see equ.~(\ref{beta_def}). }
\end{figure}

\section{Elliptic Flow}

\begin{figure}[t]
  \includegraphics[width=.45\textwidth]{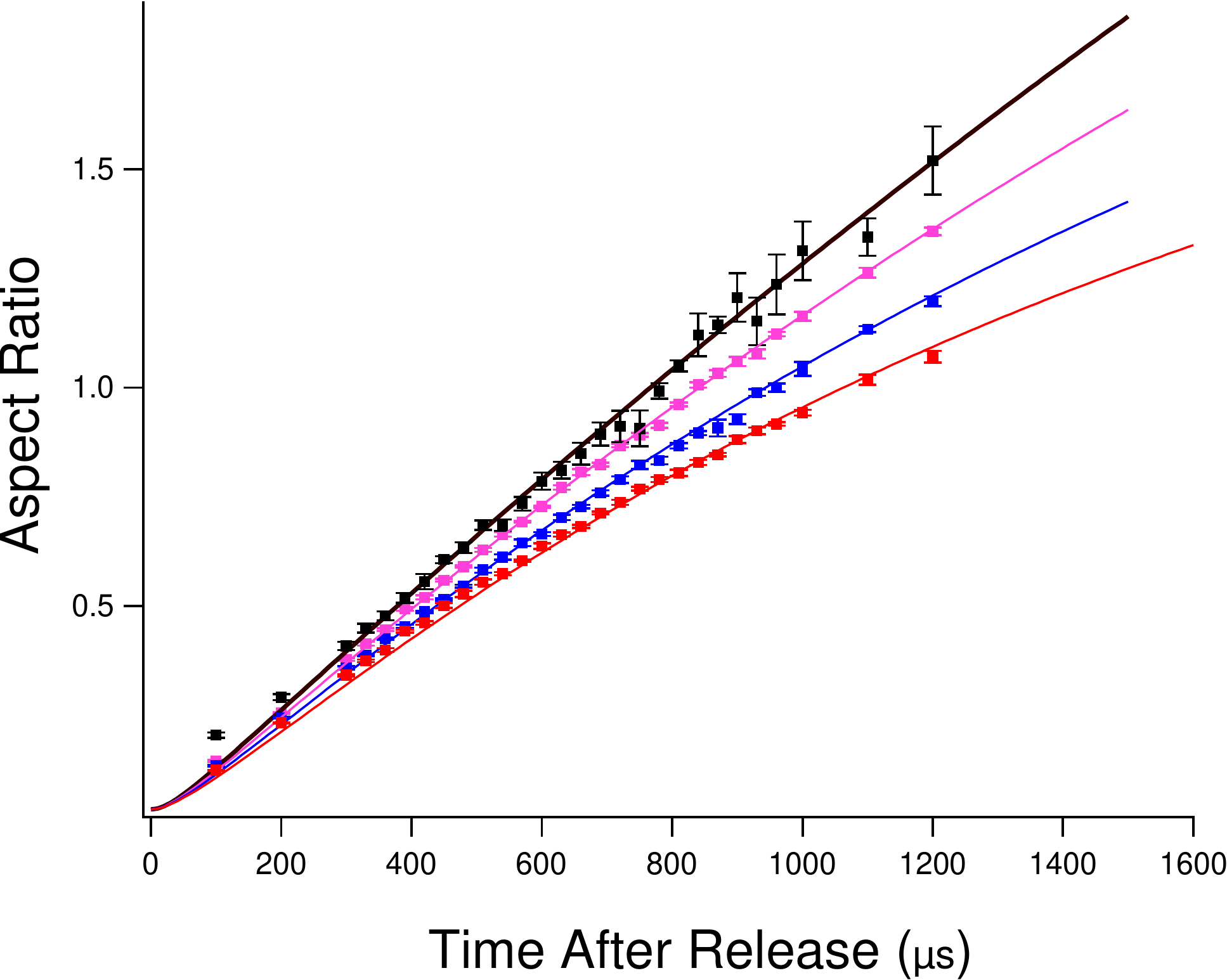}
  \caption{\label{fig_A_R_exp}
Data for the aspect ratio versus time, from \cite{Cao:2010}: Top Black, 
$E=0.6\,E_F$; Pink,  $E=2.3\,E_F$; Blue, $E=3.3\,E_F$ ; Bottom Red, 
$E=4.6\,E_F$. Solid curves: Hydrodynamic theory with the viscosity 
as the fit parameter.}
\end{figure}

 The first experiment that demonstrated nearly perfect fluidity in 
the dilute Fermi gas was the observation of elliptic flow by 
O'Hara et al.~\cite{OHara:2002}. The experiment involves releasing 
the Fermi gas from a deformed, cylindrically symmetric, trap. The 
density evolves as
\be
\label{n_scale} 
 n(x_\perp,x_z,t) = \frac{1}{b_\perp^2(t)b_z(t)}
  n_0\left(x_\perp b_\perp(t),x_z b_z(t)\right)\, , 
\ee
where $x_\perp,x_z$ are the transverse and longitudinal coordinate, 
$b_\perp(t),b_z(t)$ are scale factors, and $n_0(x_\perp,x_z)$ is 
the equilibrium density of the trapped system. The initial 
system is strongly deformed, $A_R(0)=[\langle x_\perp^2\rangle/
\langle x_z^2\rangle]^{1/2}\ll 1$. Hydrodynamic evolution 
converts the large transverse pressure gradient into transverse
flow. As a consequence the aspect ratio $A_R(t)$ grows with 
time and eventually becomes larger than one, see
Figs.~\ref{fig_A_R},\ref{fig_A_R_exp}. 

 Viscosity slows down the transverse expansion of the system.
In order to quantify the effect of shear viscosity we have to 
solve the Navier-Stokes equation for the expanding cloud
\cite{Schaefer:2009px,Schafer:2010dv}. The Navier-Stokes equation 
is 
\be
\label{ns_force}
m\left(\frac{\partial}{\partial t} +\vec{v}\cdot\vec{\nabla}\right)v_i
  =f_i + \frac{\nabla_j (\eta\,\sigma_{ij})}{n}\, , 
\label{eq:force}
\ee
where $f_i=(\nabla_i P)/m$ is the force. With the help of the 
Navier-Stokes equation the energy equation can be written as 
\bea
\label{q_force}
\lefteqn{ \left(\frac{\partial}{\partial t} + \mathbf{v}\cdot\nabla +
    \frac{2}{3}\left(\vec{\nabla}\cdot\vec{v}\right)\right)f_i }
 && \nonumber \\
 && \mbox{}  + (\nabla_i v_j)f_j 
-\frac{5}{3}\left(\nabla_i\nabla_j v_j\right)\frac{P}{n} 
  =  -\frac{2}{3}\frac{\nabla_i\,\dot{q}}{n},
\eea
where $\dot{q}=\frac{\eta}{2}(\sigma_{ij})^2$ is the heating rate. 
In general these equations have to be solved numerically on a finite
grid. In \cite{Schafer:2010dv} we showed that under certain assumptions 
a very accurate scaling solution can be derived. We will assume that 
the local shear viscosity is proportional to the density, $\eta = 
\alpha_n n$, where $\alpha_n$ is a constant. We will also assume that 
the systems remains isothermal and that heat conductivity is not 
important.

The basic idea is to postulate that the velocity field and the force are 
linear in the coordinates. If the velocity is linear and $\eta\sim n$ 
then all terms in equ.~(\ref{ns_force}) are linear in $x_i$. Also, 
equ.~(\ref{q_force}) is independent of the pressure and all
the remaining terms are linear in $x_i$. We write $f_i=a_i x_i$,
$v_i=\alpha_i x_i$ (no sum over $i$) and use the scaling ansatz 
(\ref{n_scale}) for the density. The continuity equation requires
$\alpha_i=\dot{b}_i/b_i$. The scale parameters $a_i$ and $b_i$ 
are determined by the coupled equations
\bea
\label{ns_for_1} 
\frac{\ddot b_\perp}{b_\perp}  &=&  a_\perp
   -  \frac{2\beta\omega_\perp}{b^2_\perp}
      \left( \frac{\dot b_\perp}{b_\perp} 
                - \frac{\dot b_x}{b_x} \right)\, ,  \\
\label{ns_for_2}
\frac{\ddot b_z}{b_z}  &=& a_z
   +  \frac{4\beta\lambda\omega_z}{b^2_z}
      \left( \frac{\dot b_\perp}{b_\perp} 
                - \frac{\dot b_z}{b_z} \right)\, ,\\
\label{ns_for_3}
\dot{a}_\perp  &=& 
 \!\!\!  -\frac{2}{3}\,a_\perp
   \left(5\,\frac{\dot{b}_\perp}{b_\perp} + \frac{\dot{b}_z}{b_z}\right)
  \nonumber \\
 && \hspace{0.75cm}\mbox{}
 + \frac{8\beta\omega_\perp^2}{3b_\perp}
  \left(\frac{\dot{b}_\perp}{b_\perp} - \frac{\dot{b}_z}{b_z}\right)^2
  \, , \\
\label{ns_for_4}
 \dot{a}_z  &=& 
  \!\!\! -\frac{2}{3}\,a_z
   \left(4\,\frac{\dot{b}_z}{b_z} + 2\, \frac{\dot{b}_\perp}{b_\perp}\right)
  \nonumber \\
 && \hspace{0.75cm}\mbox{}
 + \frac{8\beta\lambda\omega_z}{3b_z^2}
  \left(\frac{\dot{b}_\perp}{b_\perp} - \frac{\dot{b}_z}{b_z}\right)^2
  \, ,  
\eea
where $\omega_\perp,\omega_z$ are the oscillator frequencies of the 
harmonic confinement potential (before the gas is released). The 
parameter $\beta$ is defined by 
\be 
\label{beta_def}
\beta = \frac{\alpha_n}{(3N\lambda)^{1/3}}
        \frac{1}{(E_0/E_F)} \, ,
\ee
where $N$ is the number of atoms, $\lambda=A_R(0)$ the initial 
aspect ratio, and $E_0/E_F$ the initial energy in units of 
$E_F=(3N\lambda)^{1/3}N\omega_\perp$. The initial conditions are 
$b_\perp(0)=b_z(0)=1$, $\dot{b}_\perp(0)=\dot{b}_z(0)=0$, and 
$a_\perp(0)=\omega_\perp^2$, $a_z(0)=\omega_z^2$.

 Dissipative effects fall into two categories. The terms proportional
to $\beta$ in equ.~(\ref{ns_for_1},\ref{ns_for_2}) correspond to 
friction -- shear viscosity slows down the expansion in the transverse
direction. The dissipative terms in equ.~(\ref{ns_for_3},\ref{ns_for_4})
describe reheating -- shear viscosity converts some kinetic energy 
to heat which increases the pressure and eventually re-accelerates
the system. The interplay between these two effects can be seen 
in Fig.~\ref{fig_A_R}. As expected, friction slows down the growth 
of $A_R(t)$ as compared to the ideal evolution. Reheating reduces
this effect by about a factor of 2. We also note that the solution 
of the equ.(\ref{ns_for_1}-\ref{ns_for_4}) agrees with very well 
with numerical solutions on a three dimensional grid. 

 In order to make comparisons with data we have to take into account
that $\alpha_n$ is not a constant. In \cite{Schaefer:2009px,Schafer:2010dv} 
we argued that $\alpha_n$ in equ.~(\ref{beta_def}) should be interpreted
as the trap average of the local ratio of shear viscosity over 
density, 
\be  
\label{alpha_av}
 \langle \alpha_n\rangle = \frac{1}{N} 
   \int d^3x\, \alpha_n\!\left(\frac{mT}{n_0(x)^{2/3}}\right)n_0(x)\, .
\ee
In the dilute corona of the cloud the local shear viscosity is
independent of the density and equ.~(\ref{alpha_av}) is not well
defined. This problem can be addressed by taking into account that 
the viscous stresses relax to the Navier-Stokes value on a time 
scale $\tau$ that becomes large as the density goes to zero, see
equ.~(\ref{eta_w}). A relaxation model for $\langle\alpha_n\rangle$
was studied in \cite{Schaefer:2009px}. An even simpler model can 
be constructed based on the assumption that the shear viscosity 
relaxes to its equilibrium value at the center of the trap and is 
proportional to the density in the dilute corona. This implies that 
$\eta(x)=\eta(0)(n(x)/n(0))$. This parametrization agrees with the 
relaxation model at the 30\% level. It was used by Cao et 
al.~\cite{Cao:2010} to analyze the data shown in Fig.~\ref{fig_A_R_exp}. 
The hydrodynamic curves shown in Fig.~\ref{fig_A_R_exp} were obtained 
with $\eta=\eta_0 (mT)^{3/2}$ and $\eta_0=0.33$. This agrees quite 
well with the prediction of kinetic theory $\eta_0=15/(32\sqrt{\pi})
\simeq 0.26$, see equ.~(\ref{eta_0}).

\begin{figure}[t]
  \includegraphics[width=.47\textwidth]{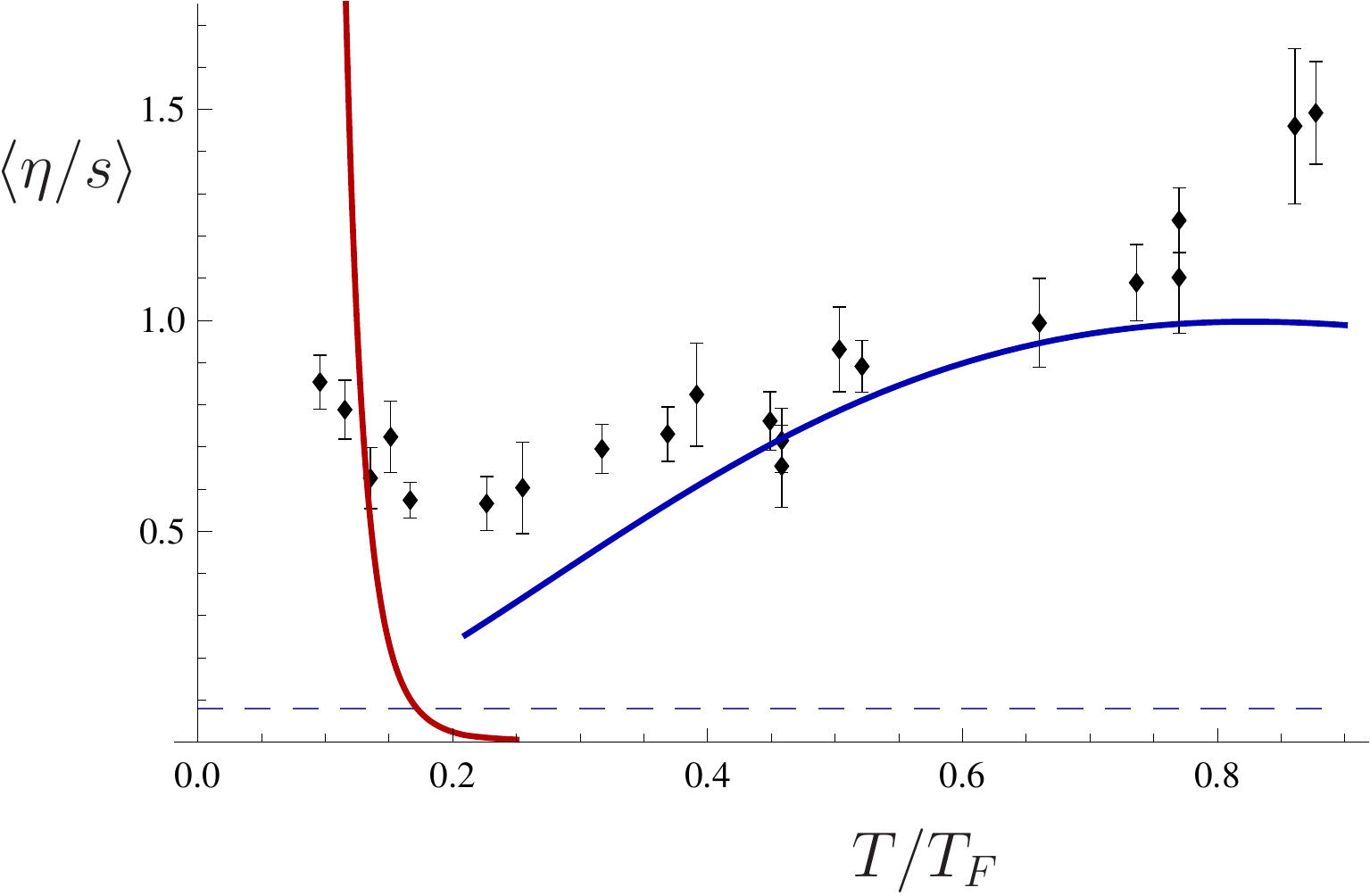}
  \caption{\label{fig_coll}
Trap average $\langle\alpha_s\rangle = \langle \eta/s\rangle$ 
extracted from the damping of the radial breathing mode. The
data points were obtained using equ.~(\ref{alpha_n_data}) to 
analyze the data published by Kinast et al.~\cite{Kinast:2004b}.
The thermodynamic quantities $(S/N)$ and $E_0/E_F$ were taken
from \cite{Luo:2008}. The solid red and blue lines show the 
expected low and high temperature limits. Both theory curves
include relaxation time effects.}
\end{figure}

\section{Collective Modes}
\label{sec_coll}

 At temperatures $T<T_F$ dissipative effects on elliptic flow 
are very small. In this regime accurate measurements of the shear 
viscosity can be obtained by analyzing the damping of collective 
modes \cite{Kinast:2004b,Schafer:2007pr,Turlapov:2007}. The 
analysis presented in the previous section can easily be 
extended to that case. A solution that describes a radial 
breathing mode is given by
\be 
\label{damp_cos}
 b_\perp(t) = 1 + a_\perp \cos(\omega t) 
   \exp\left(-\beta\omega_\perp t\right)\, . 
\ee
where $a_\perp\ll 1$ is the amplitude, $\omega=(10/3)^{1/2}
\omega_\perp$ is the frequency, and $\beta$ is the parameter 
defined in equ.~(\ref{beta_def}). The experimentally measured damping 
rate $\Gamma$ can be used to estimate $\langle\alpha_n\rangle$. 
We find \cite{Schaefer:2009px}
\be 
\label{alpha_n_data}
\langle \alpha_n \rangle = (3\lambda N)^{1/3}
   \left(\frac{\Gamma}{\omega_\perp}\right) 
   \left(\frac{E_0}{E_F}\right)\, . 
\ee
In order to compare with the heavy ion data and the proposed string 
theory bound it is interesting to convert the ratio $\eta/n$ to the 
ratio $\eta/s$ of shear viscosity to entropy density. This can be 
done using measurements of entropy per particle published in 
\cite{Luo:2008}. In Fig.~\ref{fig_coll} we show an analysis of the 
collective mode data obtained by Kinast et al.~\cite{Kinast:2004b} 
using equ.~(\ref{alpha_n_data}). The solid lines show the prediction
of kinetic theory in the limits $T\ll T_F$ and $T\gg T_F$. We observe
that $\eta/s$ is consistent with kinetic theory for $T\gsim 0.5 T_F$.
We also find that in this regime the data from collective modes are 
in agreement with the results based on the elliptic flow data 
\cite{Cao:2010}. The data do not show the expected behavior at 
very low temperature. We should note, however, that for $T\ll T_F$
the mean free path is very large and dissipative hydrodynamics is
not applicable. 

 We observe that $\eta/s$ reaches a minimum close to the phase transition 
temperature, and that the minimum value is $\eta/s\simeq 0.5$ (in units 
of $\hbar =k_B=1$). We emphasize that this result refers to trap averaged 
quantities. We expect that improved data will allow us to determine the 
local value of the ratio $\eta/s$.

\section{Conclusions and outlook}

 There are a number of issues that need to be addressed before 
an accurate value of $\eta/s$ with fully controlled errors can 
be obtained. The most important of these is a better description 
of the transition from nearly perfect fluid dynamics in the 
center of the cloud to kinetic behavior in the dilute corona. 

 On the the theoretical side we need to develop tools that 
will allow us to perform calculations of transport properties
in the strongly coupled regime. Some steps in this direction 
have been taken. Taylor and Randeria derived sum rules for the 
viscosity spectral function \cite{Taylor:2010ju}. Enss, Haussmann 
and Zwerger developed a resummed diagrammatic scheme that respects 
the sum rule and reproduces the kinetic limit \cite{Enss:2010qh}. 
There have also been some attempts at extending the AdS/CFT 
correspondence to non-relativistic conformally invariant
theories, see \cite{Balasubramanian:2008dm,Son:2008ye}.

 Finally, we would like to understand how nearly perfect fluidity 
arises in different physical systems. In the case of the quark gluon 
plasma the question is whether nearly perfect fluidity is caused by 
strong interactions between well-defined quark and gluon quasi-particles, 
or whether the quasi-particle picture breaks down completely and the 
low energy description involves non-local degrees of freedom, as in 
the AdS/CFT correspondence. In the case of the dilute Fermi gas 
we would like to understand whether momentum transport is governed
by fermionic quasi-particles, by collective modes (like the roton 
in liquid Helium), or whether there is no quasi-particle description 
at all. This question is difficult to address experimentally, but 
it can be studied numerically, by computing the viscosity spectral 
function. These calculations are still in their infancy, but 
we expect significant progress in the near future. 


\begin{theacknowledgments}
  This work was supported in parts by the US Department of Energy 
grant DE-FG02-03ER41260. Part of the work reported here was done 
in collaboration with M.~Braby, C.~Chafin, J.~Chao and J.~Thomas.
\end{theacknowledgments}

\end{document}